\documentclass[letterpaper, 10 pt, conference]{ieeeconf}  

\IEEEoverridecommandlockouts                              

\usepackage{multicol}
\usepackage[bookmarks=true]{hyperref}
\usepackage[table]{xcolor}
\usepackage[noend]{algpseudocode}
\usepackage[ruled,vlined]{algorithm2e}
\usepackage{graphicx}
\usepackage{grffile}
\usepackage{times} 
\usepackage{amsmath} 
\usepackage{amssymb}  
\usepackage{multirow}
\usepackage{diagbox}
\usepackage{svg}
\usepackage{caption}
\usepackage{tikz}
\usepackage{adjustbox}
\usepackage{subcaption}
\usepackage{booktabs}

\newcommand{\ourmethod}{C-Uniform}

\newcommand{\newourmethod}{Neural C-Uniform}
\newcommand{\newsampourmethod}{CU-MPPI}
\newcommand{\newsampourmethodlog}{CU-LogMPPI}
\newcommand{\statevec}{\mathbf{x}}

\title{\LARGE \bf
An Unsupervised C-Uniform Trajectory Sampler with Applications to Model Predictive Path Integral Control
}

\author{O. Goktug Poyrazoglu$^{\dagger}$, Rahul Moorthy$^{\dagger}$, Yukang Cao, William Chastek, and Volkan Isler
\thanks{${}^{\dagger}$: Equal contribution {\tt \{poyra002, mahes092\}@umn.edu}. The authors are with the Robotics, Sensing and Networks Laboratory (RSN) at the University of Minnesota.}
}

\begin{document}

\maketitle
\thispagestyle{empty}
\pagestyle{empty}

\begin{abstract}
Sampling-based model predictive controllers generate trajectories by sampling control inputs from a fixed, simple distribution such as the normal or uniform distributions. This sampling method yields trajectory samples that are tightly clustered around a mean trajectory. This clustering behavior in turn,  limits the exploration capability of the controller and reduces the likelihood of finding feasible solutions in complex environments. Recent work has attempted to address this problem by either reshaping the resulting trajectory distribution or increasing the sample entropy to enhance diversity and promote exploration. 
In our recent work, we introduced the concept of  \ourmethod{} trajectory generation~\cite{poyrazoglu2024cuniformtrajectorysamplingfast} which allows the computation of control input probabilities to generate trajectories that sample the configuration space uniformly. In this work, we first address the main limitation of this method: lack of scalability due to computational complexity. 
We introduce 
\newourmethod{}, an unsupervised \ourmethod{} trajectory sampler that mitigates scalability issues by computing control input probabilities without relying on a discretized configuration space. Experiments show that \newourmethod{} achieves a similar uniformity ratio to the original \ourmethod{} approach and  
generates trajectories over a longer time horizon while preserving uniformity.
Next, we present \newsampourmethod{}, which integrates \newourmethod{} sampling into existing MPPI variants. 
We analyze the performance of \newsampourmethod{} in simulation and real-world experiments. Our results indicate that in settings where the optimal solution has high curvature, \newsampourmethod{} leads to drastic improvements in performance.

\end{abstract}


\section{Introduction}

\begin{figure}[t]
    \centering
    \begin{subfigure}{0.2\textwidth}
        \includegraphics[width=\textwidth]{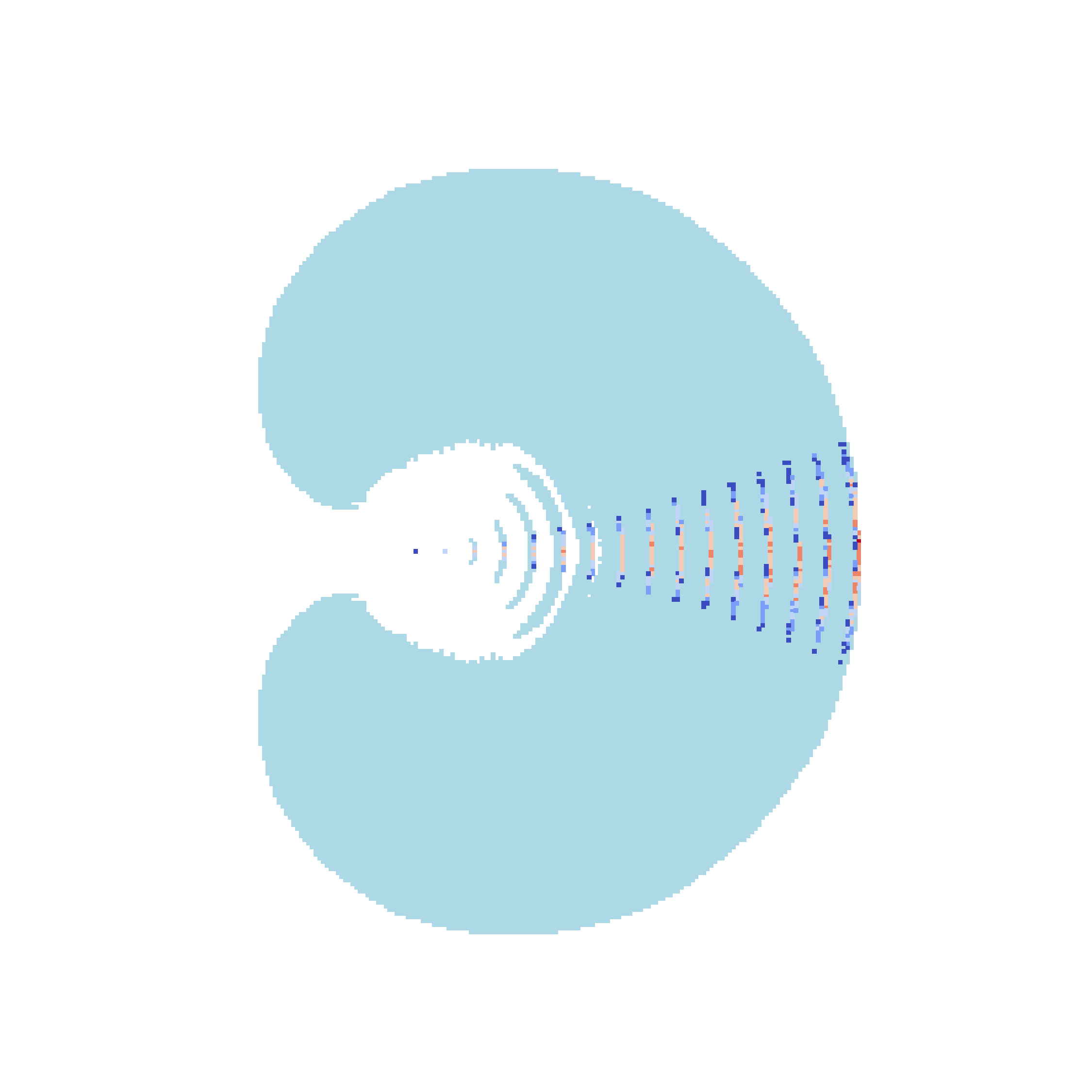}
        \caption{MPPI}
    \end{subfigure}
    \hfill
    \begin{subfigure}{0.2\textwidth}
        \includegraphics[width=\textwidth]{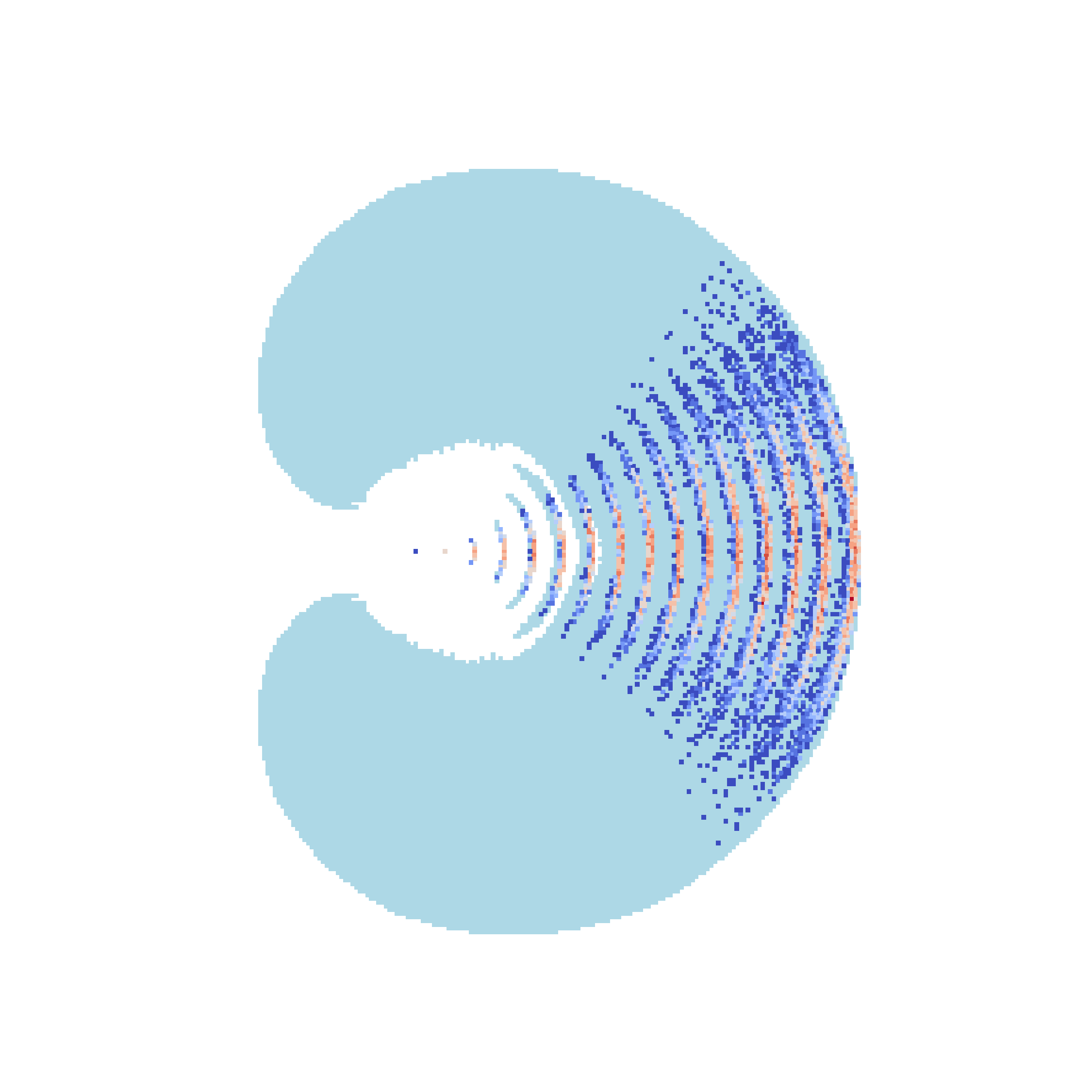}
        \caption{Log-MPPI}
    \end{subfigure}
    \hfill
    \begin{subfigure}{0.2\textwidth}
        \includegraphics[width=\textwidth]{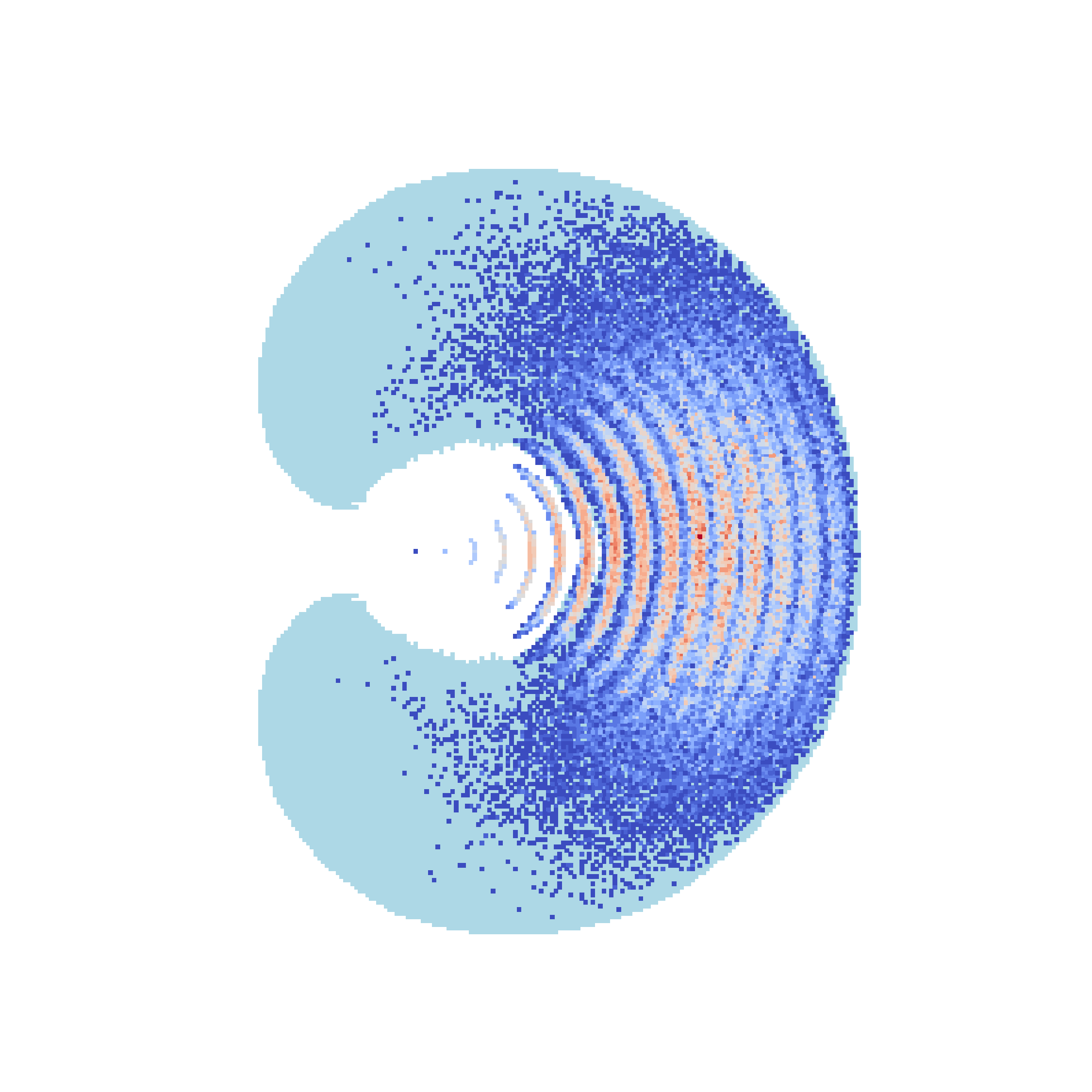}
        \caption{Neural C-Uniform}
    \end{subfigure}
    \hfill
    \begin{subfigure}{0.2\textwidth}
        \includegraphics[width=\textwidth]{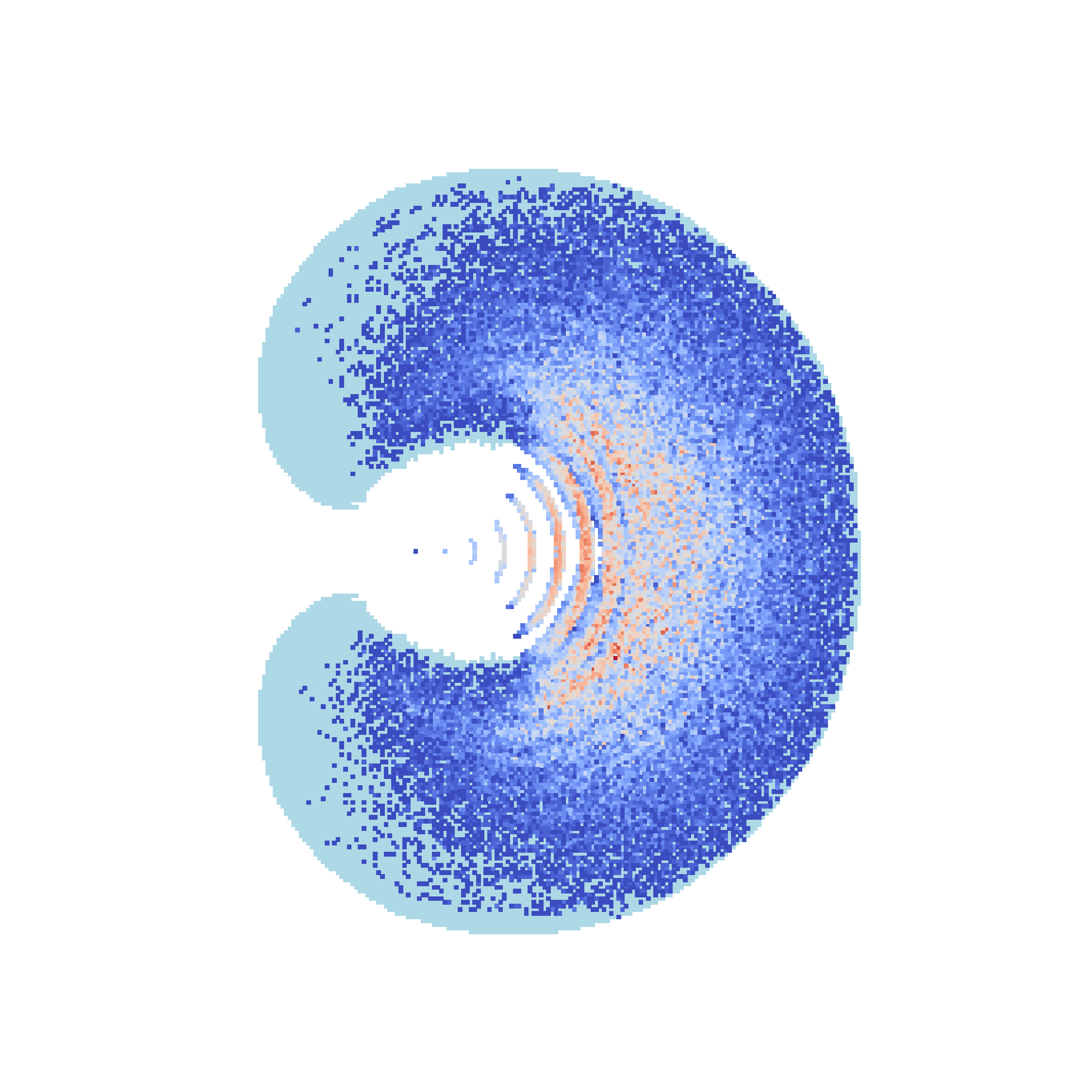}
        \caption{\ourmethod}
    \end{subfigure}

    \caption{Comparison of trajectory samples.
     The robot's configuration space is $(x,y,\theta)$. The forward speed is constant. The steering angle is directly controllable within 
     $[-30, +30]$ degrees/second. 
    The light blue area is the portion of the plane reachable within 3s. The dark blue part is the region visited by 10K trajectories generated by each approach. The color transition illustrates coverage in $\theta$ (dark blue indicates lowest to red the highest). \ourmethod{} and \newourmethod{} show high diversity in all three dimensions of the configuration space compared to MPPI and Log-MPPI.}
    \label{fig:trajectory_comparison}
\end{figure}

Sampling-based model predictive controllers generate ``minimum cost" trajectories using a set of trajectory samples to achieve objectives such as arriving at a goal location while avoiding obstacles and adhering to motion constraints. They have been used in various robotics applications including autonomous driving~\cite{williams2018information}, manipulation~\cite{lambert2020stein}, and drone navigation~\cite{minavrik2024model}. 
In order to generate random trajectories which are also kinematically valid, existing methods sample control inputs using a simple distribution such as the normal distribution. The system model is then used to propagate the state using these random inputs. However, as shown in Fig.~\ref{fig:trajectory_comparison}, these sampling strategy generally yield samples that are clustered
around a mean trajectory which limits the exploration capacity of the controller and reduces the likelihood of finding feasible solutions.

Recent works have addressed this issue of insufficient exploration by modifying control input sampling distribution to promote higher trajectory sample diversity and enhanced exploration. For example, log-MPPI~\cite{mohamed_autonomous_2022} introduced a new sampling distribution, normal-log-normal, to flatten the resulting trajectory distribution. Even though this approach enhances exploration during trajectory sampling, the exploration is still local which can be problematic in high-curvature or multimodal settings. 

 Rather than modifying the final trajectory distribution,  an alternative is to focus on finding a better nominal trajectory to initialize the sampling mechanism to avoid the mode collapse. This is achieved by generating a set of proposals and optimizing them to locate trajectories around lower-cost areas so that subsequent sampling can be performed around those regions. Recently introduced Stein Variational Guided Model Predictive Path Integral Control (SVG-MPPI)~\cite{honda2024steinvariationalguidedmodel} integrates Stein Variational Gradient Descent (SVGD) to find a good nominal trajectory. It iteratively refines a set of trajectory samples by pushing them toward lower-cost regions. This mode-seeking behavior helps to concentrate the trajectory sampling around low-cost regions in the cost landscape. However, the efficiency of this process heavily depends on the quality of the initial sample set. For example, if the starting trajectories do not adequately cover the C-space, the refinement process may require many iterations with the added computational cost, or the gradient may provide limited information that causes the solution to get trapped in a local minimum.

In our previous work, we presented the \ourmethod{}~\cite{poyrazoglu2024cuniformtrajectorysamplingfast} trajectory sampling method that computes control input probabilities to generate trajectories that uniformly sample the configuration space (C-space). In other words, \ourmethod{} provides a systematic and unified approach to exploration. However, it relies on discretization of the configuration space to build a flow network to compute the optimal flow which is costly both in computation time and space. 

In this paper, we address these limitations using an unsupervised learning approach and present the \emph{\newourmethod{} trajectory sampling method}, in which a neural network is trained to map the state to control input probabilities that lead to C-Uniform trajectories.  
This approach eliminates the need for discretization and enables the generation of trajectories for longer horizons while maintaining uniformity (Fig.~\ref{fig:trajectory_comparison}). 
Our second contribution is 
a new variant of MPPI, \newsampourmethod{}, that leverages trajectories from \newourmethod{} to increase the chances of finding a better nominal trajectory by covering the C-space uniformly and avoiding the dependence on the gradient. 

In summary, the contributions of our work are:
\begin{itemize}
    \item  We present \newourmethod{} trajectory sampler, which uses entropy maximization formulation to generate trajectories that are uniform in the configuration space (Sec.~\ref{sec:neural_cuniform}). 
    \item  We present \newsampourmethod{}, a new sampling-based model predictive controller that utilizes \newourmethod{} trajectories to enhance exploration. By ensuring broad coverage of the C-space, our method increases the likelihood of finding the global minimum regions while reducing dependence on gradient-based refinements (Sec.~\ref{sec:cu_mppi}).
    \item We perform experimental validation through real-world and simulation experiments to assess the advantages of having a diverse trajectory sampling strategy and its effectiveness in sampling-based model predictive controller settings (Sec.~\ref{sec:experiments}).
\end{itemize}

The results indicate that the notion of C-Uniformity provides a systematic trade-off between exploration and gradient-seeking (exploitation) for MPPI-based methods. We start with an overview of related work.

\section{Related Work}

Existing approaches addressing the need for diversity in trajectory sampling for sampling-based planners/controllers can be broadly categorized into two main groups.
\vspace{-3pt}
\subsection{Trajectory Sampling}
Trajectory sampling arises in a wide range of research domains, such as stochastic processes \cite{kalakrishnan2011stomp}, control theory \cite{kazim2024recent}, motion planning \cite{orthey_sampling-based_2024}, and reinforcement learning \cite{ota_trajectory_2019}. It is often employed to study system behavior under uncertainty to get insights into the probabilistic dynamics of stochastic processes \cite{kalakrishnan2011stomp}. In control theory, it is used to design and analyze control inputs that navigate systems along desired objectives while satisfying the constraints \cite{williams2018information}; on the motion planning side, it is used to determine a path or a trajectory to guide systems, while similarly adhering to constraints \cite{lavalle_planning_2006}. Even though there is extensive research on trajectory sampling, it turns out that determining controls for the trajectory distribution remains a relatively less explored area. The two works closest to this context are C-Uniform trajectory sampling and sample-based MPC. Thomas et al~\cite{power2024learning} proposed sampling-based MPC to generate control inputs for collision-free paths using a normalizing flow as a sampling distribution. In our previous work~\cite{poyrazoglu2024cuniformtrajectorysamplingfast}, we proposed C-Uniform trajectory sampling, which concerns uniformly sampling the set of valid configuration space using robot inputs to maintain the desired trajectory distribution over time. 



\subsection{Sampling-based Model Predictive Control}

In recent years, with the enhancement of parallel computing power, sampling-based model predictive control methods (SBMPC) have increased in popularity \cite{abughalieh2019survey}. Pioneer work called Model Predictive Path Integral (MPPI) control combines the path integral theory and MPC formulation \cite{williams2018information}. In that work, control inputs are sampled using a Gaussian to generate a tractable and controllable trajectory distribution around a nominal trajectory. However, the Gaussian assumption leads to vital problems in changing environment settings. Researchers have addressed this issue in several ways. In~\cite{yin2022trajectory}, covariance steering theory is used to shift the final shape of trajectory distribution. Similarly, log-MPPI~\cite{mohamed_autonomous_2022}, uses normal-log-normal distribution to flatten the resulting sampling distribution or adding bias to the cost distribution allows to have arbitrary sampling distributions~\cite{trevisan2024biased}. Moreover, some methods use adaptive importance sampling~\cite{asmar2023model} to shift the solution to the lower-cost regions. Alternatively, other methods tend to move the nominal cost to areas with low-cost by solving reverse Kullback-Leibler divergence to find a mode of the cost distribution~\cite{kobayashi2022real}. Comparably, Stein Variational Gradient Descent is also used for understanding cost distribution~\cite{lambert2020stein} or guiding the MPPI trajectories to low-cost regions by modifying both the nominal trajectory and the covariances~\cite{honda2024steinvariationalguidedmodel}.

\section{Preliminaries}

In this section, we summarize important definitions, dynamic models, and concepts required to develop our approach. After these definitions, we explain what \ourmethod{} trajectories are in Sec.~\ref{sec:cuniform_traj_samp}.



We consider a robotic system with a state vector $\statevec \in \mathcal{X} \subseteq \mathcal{R}^p$, and a control input vector $\mathbf{u} \in \mathcal{U} \subseteq \mathcal{R}^q$ with the known discrete dynamic model $\statevec_{t+1} = \mathbf{F}(\statevec_t, \mathbf{u}_t)$. We use a kinematic bicycle car model as our vehicle model where $\statevec = [x,y, \psi] \in \mathbb{R}^3 $ is the vehicle's state, and $\mathbf{u} = [v, \delta]$ is the steering angle and the linear velocity $v$ is constant. The state transition is defined using Equation \ref{eq;kinematic_bicycle_model}.

\vspace{-10pt}
\begin{align}   \label{eq;kinematic_bicycle_model} 
    \statevec_{t+1} &= \mathbf{F}(\statevec_t, u) = 
    \begin{bmatrix}
        x_t \\
        y_t \\
        \psi_t
    \end{bmatrix} \Delta_t
    =
    \begin{bmatrix}
        v\cos\psi_t \\
        v\sin\psi_t \\
        \frac{v}{L_{wb}}\tan(\delta_t)
    \end{bmatrix} \Delta_t
\end{align}
where $\Delta_t$ is the time discretization and $L_{wb}$ is the vehicle wheelbase length.

We also define finite horizon trajectories with horizon $T$ as $\tau = \mathbf{F}(\statevec_0, \mathbf{U})$, where $\statevec_0$ is the initial state, and $\mathbf{U} =(\mathbf{u}_0, \mathbf{u}_1, \dots, \mathbf{u}_{T-1})$ is the control sequence. The resulting trajectory $\tau$ is generated by recursively applying $F$ with corresponding state $\statevec_t$ and control $\mathbf{u}_t$ pairs. It is assumed that the dynamic model $F$ satisfies the Lipschitz continuity, which means the state propagation with this model is predictable.

\subsection{C-Uniform Trajectory Sampling} \label{sec:cuniform_traj_samp}

We provide key concepts of the \ourmethod{} trajectory sampling. For more details, we direct the reader to prior research~\cite{poyrazoglu2024cuniformtrajectorysamplingfast}. We first define a \textit{Level Set} $L_t$ as the following equation:
\vspace{-3pt}
\begin{equation}
L_t = \{ \statevec_t \in \mathcal{X} \ | \ \exists \mathbf{U}  =(\mathbf{u}_0, \mathbf{u}_1, \dots, \mathbf{u}_{t-1}) \text{ s.t. } \statevec_t = F(\statevec_0, \mathbf{U}) \}.
\end{equation}
where the set of all states $\statevec_t \in \mathcal{X}$ such that there is a control sequence $\mathbf{U}$ of length $t$ with $\statevec = F(\statevec_0, \mathbf{U})$. We also need to note that these level sets are disjoint, meaning if a state $\statevec_i$ is already covered by some $L_i$, we discard that state that is also reachable in any later level set $L_j$, where $j > i$. Lastly, we define $L_{D,t}$ as the discretized version of $L_{t}$ where each representative state $\statevec_{D,t} \in L_{D,t}$ is found by $L_{t}/\delta$ where $\delta$ defines a small measurable uniform region of \ourmethod. We use the Lebesgue measure ($\mu$) to quantify the size of these regions. Then, the uniform probability of representatives for each level set is defined as $P(\statevec_{t} \in \delta) = \mu(\delta)/\mu(L_t)$, where $\statevec_{t}$ is a state in the uniformity cell. 

The probabilities associated with level sets are computed recursively for each level set as follows. 
\vspace{-3pt}
\begin{equation}
p(\statevec_{t+1}) = \sum_{\statevec_t \in L_t} \sum_{\mathbf{u_i}: \statevec_{t+1}= F(\statevec_t, \mathbf{u_i}) } p(u_i|\statevec_t)p(\statevec_t)
\label{eqn:lprobs}
\end{equation}
where $\statevec_{t+1}$ is the state in the next level set, and the control inputs $\mathbf{u}_i$ are the ones that reach the state $\statevec_{t+1}$ by propagating the current state $\statevec_t$. Additionally, we also discretize the action space $\mathcal{U}$ into a set of distinct actions $\mathcal{U} = \{u_0,u_1,\dots,u_N\}$ and we define a probability mass function (pmf) over it. The pmf is denoted as $p(\mathcal{U}|\statevec)$, where $\sum_{i=0}^N p(u_i|\statevec) = 1$, and $p(u_i|\statevec) \geq 0$. 

By introducing the Eq.~\ref{eqn:lprobs}, we can state a similar problem formulation for the \newourmethod{} trajectory sampling, as defined in~\cite{poyrazoglu2024cuniformtrajectorysamplingfast}: 

Given an initial state $\statevec_0$ and a system's dynamic model $F$ determine control action probabilities $p(\mathcal{U}|\statevec)$ for each state such that the probability distribution associated with each level set is uniform.

\section{Unsupervised \ourmethod{} Trajectory Sampler}
\label{sec:neural_cuniform}

In this section, we introduce our \newourmethod{} trajectory sampling approach. We first present an entropy maximization formulation of  \newourmethod{} to generate the probability distribution of control inputs $p(\mathcal{U}|\statevec)$ that satisfies the Eq.~\ref{eqn:lprobs}. In particular, given a state $\statevec$, we generate a probability distribution of control inputs $p(\mathcal{U}|\statevec)$ which uniformly samples all level sets.

\subsection{Entropy Maximization Formulation}
Uniform distribution leads to maximum entropy which is unique among all probability distributions defined over a domain~\cite{guiasu1985principle}. 
Hence, we formulate \newourmethod{} as an iterative level set entropy maximization problem to learn generating $p_\theta(\mathcal{U}|\statevec)$ parameterized by $\theta$ for each state $\statevec_t$ in $L_{t}$ and resulting in $\statevec_{t+1} =  F(\statevec_t, \mathbf{u_t})$ in $L_{t+1}$ to maximize for entropy $\mathcal{H}(\statevec_{t+1})$ defined by Eq.~\ref{eq:entropy_maximization}.
\vspace{-2pt}
\begin{equation}
\begin{split}
&\max_{p(\statevec_{t+1})} \mathcal{H}(\statevec_{t+1}) = \\
&\max_{p(\statevec_{t+1})} \left( - \sum_{\statevec_t} \sum_{\mathbf{u}} 
    p(\statevec_{t+1}|\mathbf{u},\statevec_t) \log p(\statevec_{t+1}|\mathbf{u},\statevec_t) \right),
\end{split}
\label{eq:entropy_maximization}
\end{equation}
where $p(\statevec_{t+1})$ shows the probability of state $\statevec_{t+1} \in L_{t+1}$ when an action $\mathbf{u}$ is taken from the $\statevec_{t} \in L_{t}$. We now describe the training procedure and network architecture of \newourmethod{} using Eq.~\ref{eq:entropy_maximization}.


\subsection{Network Architecture}
\newourmethod{} architecture is capable of learning to determine $p(\mathcal{U}|\statevec)$ for any state $\statevec$ from any level set $L_{t}$ to uniformly sample the next level set $L_{t+1}$ provided the level sets are disjoint. \newourmethod{} architecture consists of two linear layers of 256 nodes and an output layer of 45 actions representing the action distribution. The intermediate layers are applied with the ReLU activation function to capture the non-linearity of the level-set propagation over time while the output is applied with softmax to convert logits to a probability distribution.
Additionally, Batchnorm layers were added after each ReLU activation function of intermediate layers.
The input to the architecture is $\statevec \in \mathbb{R}^n$ where $n$ is the dimension of the state vector of the given dynamics system. In our case, the input is $\statevec = [x,y, \psi] \in \mathbb{R}^3 $ as shown in Eqn.~\ref{eq;kinematic_bicycle_model} where $x$, $y$ represents the positions and $\psi$ represents the heading of the vehicle which is converted to $\cos{\psi}$ and $\sin{\psi}$ to account for the periodicity. 

\begin{algorithm}
\caption{Training} \label{algo:unsupervised}
\KwIn{
    $L_{D}$: Discretized C-Uniform Level Sets;\\
    $N_t$: The number of time steps in a trajectory;\\
    $\mathcal{U}$: the set of actions;
}

$L_{D} = \{L_{D,t}\}_{t=0}^{N}$ where $L_{D,t}$ is the discretized C-Uniform Level set at time step $t$\;

\For{each time step $t = 0$ to $N-1$}{
    
    \For{each state $\statevec_{D,t} \in L_{D,t}$}{
            \For{each action $\mathbf{u} \in \mathcal{U}$}{
            $\statevec_{t+1}\leftarrow \mathbf{F}(\statevec_{D,t},\mathbf{u})$

            $Dist(\statevec_{D,t+1}) = -||\statevec_{t+1} - \statevec_{D,t+1}||_2$
            
            $p(\statevec_{D,t+1}|\mathbf{u},\statevec_{D,t}) = p_\theta(\mathbf{u}|\statevec_{D,t})\cdot e^{Dist(\statevec_{D,t+1})}$
        }
    }
        $\mathcal{H} = \sum_{\statevec_{D,t}} \sum_{\mathbf{u}} 
    p(\statevec_{t+1}|\mathbf{u},\statevec_{D,t}) \log p(\statevec_{t+1}|\mathbf{u},\statevec_{D,t})$ 
    
    $\theta = \theta - \alpha \nabla\mathcal{H}(\theta)$  
}
\end{algorithm}

\newourmethod{} architecture
is trained for entropy maximization which is defined by Eq.~\ref{eq:entropy_maximization}. In particular, $- \sum_{\statevec_t} \sum_{\mathbf{u}} 
    p(\statevec_{t+1}|\mathbf{u},\statevec_t) \log p(\statevec_{t+1}|\mathbf{u},\statevec_t) $ is used as the loss function where the negative sign is omitted to account for the gradient descent step. In order to maximize with respect to $p(\statevec_{t+1})$, we first generate next states $\statevec_{t + 1}$ using action space $\mathcal{U}$ for each discretized states $\statevec_{D,t}$ of \ourmethod\ $L_{D,t}$ and perform assignment to $\statevec_{D,t+1}$ in $L_{D, t+1}$ for calculating $p(\statevec_{t+1})$. To ensure differentiable loss, we use a soft assignment which is defined by Eq.~\ref{eq:entropy_maximization_diffrentiable}.

\vspace{-15pt}
\begin{equation}
p(\statevec_{D,t+1}|\mathbf{u},\statevec_{D,t}) \propto p_\theta(\mathbf{u}|\statevec_{D,t})\cdot e^{-||\statevec_{t+1} - \statevec_{D,t+1}||_2}
\label{eq:entropy_maximization_diffrentiable}
\end{equation}
where $x_{t+1} =  \mathbf{F}(\statevec_{D,t}, \mathbf{u})$, $p_\theta(\mathbf{u}|\statevec_{D,t})$ is the probability for an action $\mathbf{u}$ for a given state $\statevec_{D,t}$ estimated by \newourmethod{} architecture and $e^{-||\statevec_{t+1} - \statevec_{D,t+1}||_2}$ estimates p($\statevec_{D,t+1}|\statevec_{t+1}$).
The soft assignment is then used to maximize $\mathcal{H}(\statevec_{D,t+1})$ resulting in determining $p_\theta(\mathbf{u}|\statevec_{D,t})$ which uniformly samples all level sets. 
\par We use Adam optimizer with a learning rate of 0.0001 and train it for 20 epochs on a dataset of $L_D$ consisting of a time horizon of 3 seconds with 0.2 time discretization resulting in 16 level sets. Algorithm \ref{algo:unsupervised} shows the full end-to-end training pipeline of \newourmethod{} architecture where $\statevec_{D,t}$ is in $L_{D,t}$ and $\statevec_{D,t+1}$ is in $L_{D,t+1}$.

\begin{figure}[ht]
    \centering
    \subfloat[Nominal Trajectory Selection]{\includegraphics[width=0.3\linewidth,height = 3cm]{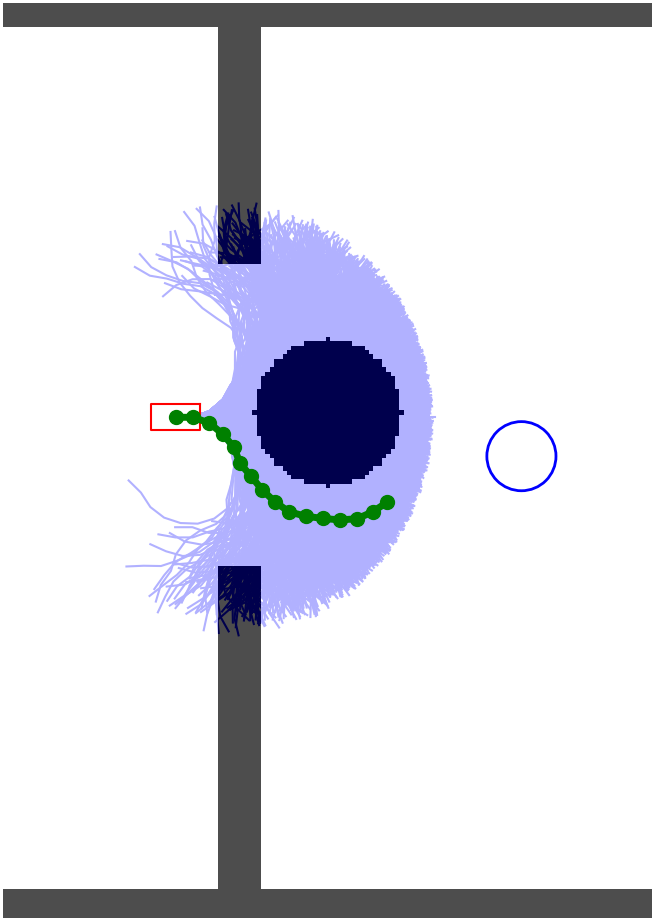}}
    \hfill
    \subfloat[Trajectory Sampling]{\includegraphics[width=0.3\linewidth,height = 3cm]{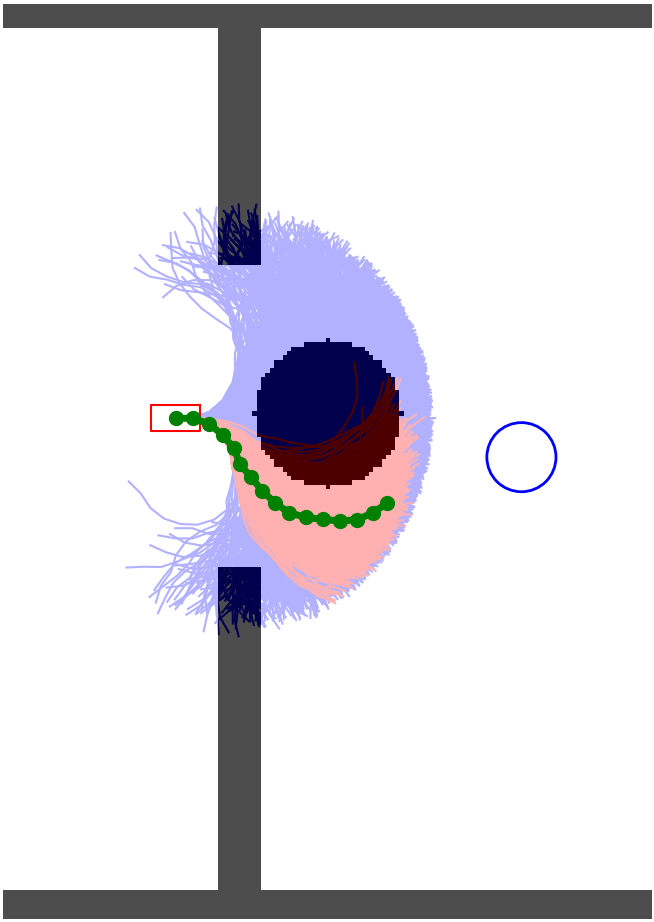}}
    \hfill
    \subfloat[Final Solution]{\includegraphics[width=0.3\linewidth,height = 3cm]{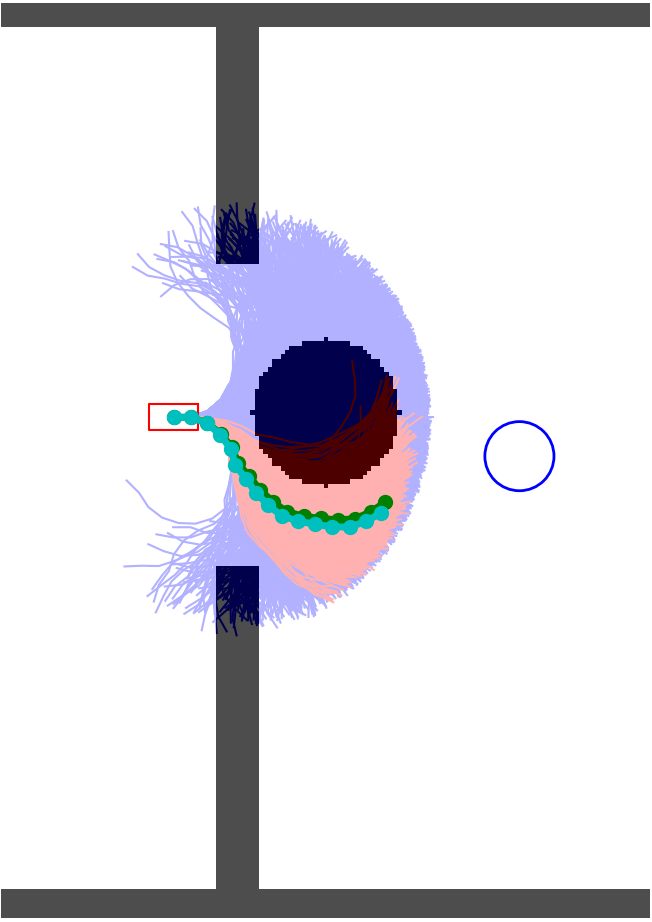}}

    \caption{\newsampourmethod{}: The \newourmethod{} trajectories represented by blue in Figure (a) are first evaluated for the cost and the trajectory with minimum cost is selected and shown in green. MPPI is initialized using the green trajectory as the nominal in Figure (b) and the MPPI-generated trajectories are shown in red. Lastly, Figure (c) shows the final trajectory generated using control signals of MPPI, which is represented by cyan. Additionally, the red rectangle represents the current vehicle configuration, and the blue circle shows the target position.}
    \label{fig:cuni_mppi_pipe}
\end{figure}

\section{C-Uniform Based Model Predictive Path Integral}
\label{sec:cu_mppi}

In this section, we introduce a new MPPI variant, \newsampourmethod{}.
We illustrate the general overview of our approach in Fig.~\ref{fig:cuni_mppi_pipe}. The method first generates a set of \newourmethod{} trajectories for the current state $\statevec_t$. Then, our approach selects the trajectory with the lowest cost from the samples set. Once the lowest-cost trajectory identified, we use it as the nominal trajectory for MPPI trajectory sampling. The MPPI algorithm solves the optimization problem in Eq.~\ref{eq:opt_u_calc} to get the final input sequence for the current time step $t$.

A key advantage of our \newourmethod{} sampling strategy is its ability to improve the likelihood of selecting a near-optimal trajectory as the number of samples increases by ensuring uniform coverage of the trajectory space. 
Let $ \mathcal{T} = \{\tau_1, \tau_2, \dots, \tau_N\}$ represent the set of $N$ sampled \newourmethod{} trajectories, each associated with a cost $J^{\tau_i} = J(\statevec_t, U)$. 

Instead of expectation minimization of the cost, as in~\cite{williams2018information}, we use direct cost minimization of the cost for the nominal trajectory selection as: 
\vspace{-5pt}
\begin{equation}
    \tau^* = \arg\min_{\tau_i \in \mathcal{T}} J^{\tau_i}.
\end{equation}

where $\tau^*$ is the minimum cost trajectory in the set $\mathcal{T}$. 

As the number of samples increases, the uniform coverage of the trajectory space improves, leading to a higher probability change to get a trajectory around low-cost regions. In cases where the cost function has a multimodal distribution, where multiple trajectories have the same minimum cost, we break ties by selecting one trajectory uniformly at random to ensure unbiased selection among optimal candidates.






\subsection{MPPI Trajectory Sampling}
    In Fig.\ref{fig:cuni_mppi_pipe}, the green trajectory represents $\tau^*$, which has the minimum cost among the sampled trajectories shown in blue. Furthermore, the action sequence $\mathbf{U}$ that generates the trajectory $\tau^* = F(\statevec_t, \mathbf{U})$ is used as the nominal control input sequence $\tilde{\mathbf{U}}$ in MPPI algorithm with a fixed covariance matrix $\Sigma$. 
    
    After the nominal input sequence is selected and fed into the MPPI algorithm. The red trajectories in Fig.~\ref{fig:cuni_mppi_pipe} show the sampled trajectories with the factorized Gaussian probability density function $q(\mathbf{V}|\tilde{\mathbf{U}},\Sigma)$. Then, the optimal action sequence by those samples is calculated by:
    \vspace{-5pt}
    \begin{equation}\label{eq:opt_u_calc} \mathbf{U}^* = \mathbb{E}_{\mathbb{Q}}[w(\mathbf{V})\mathbf{V}], \quad w(\mathbf{V}) = \frac{q^*(\mathbf{V})}{q(\mathbf{V}|\mathbf{U},\Sigma)}. \end{equation}

where, $\mathbf{V} = \mathbf{U} + \epsilon$ and $\epsilon \sim \mathcal{N}(0,\Sigma)$. In practice, Monte Carlo sampling methods are used to approximate the expectation. Further details on the derivation can be found in~\cite{williams2018information}. The optimal sequence $\mathbf{U}^*$ is shown in cyan in Fig.\ref{fig:cuni_mppi_pipe}. 

\subsection{Nominal Trajectory Selection}
In each control iteration, the method adds the MPPI optimal solution into the \newourmethod{} trajectories so that the optimal trajectory from the previous time step can also be considered for the next nominal sequence selection. This integration enhances exploitation by increasing the likelihood of staying within a low-cost region. At the same time, the inherent stochasticity of C-Uniform sampling preserves exploration. Even if MPPI provides a non-optimal solution, this exploration ability prevents premature convergence to suboptimal solutions. 

\section{Experiments}
\label{sec:experiments}
We evaluate the \newourmethod{} sampling method and \newsampourmethod{} controller by studying the following questions through experiments.

\begin{enumerate}
    \item Can the \newourmethod{} trajectory sampling method maintain uniformity over a given planning horizon, and how does the uniformity of sampled trajectories change when extrapolated to longer horizons than it was trained on? (Sec.~\ref{sec:uniformity_analysis_iros})
    \item Can the proposed controller algorithm effectively find optimal paths even when the curvature of the optimal solution is high? (Sec.~\ref{sec:open_space_exp_iros})
    \item Can the proposed controller algorithm adapt and perform reliably in dynamic and complex environments in both simulation and real-world scenarios? (Secs.~\ref{sec:simulation_exp_iros} and~\ref{sec:real_world_exp_iros}.)
\end{enumerate}

\subsection{Experimental Setup}

\textbf{Baselines:} In our simulation and real-world experiments, we compare our method against three baselines controllers: MPPI~\cite{williams2018information}, log-MPPI~\cite{mohamed_autonomous_2022}, and SVG-MPPI~\cite{honda2024steinvariationalguidedmodel}.

MPPI and log-MPPI are selected to highlight how different trajectory distributions affect the performance of various navigation tasks. MPPI only uses Gaussian samples around a nominal trajectory, and log-MPPI uses Normal-Log-Normal distribution to generate samples. We implemented both methods with a temperature parameter of $\lambda = 0.5$. We also initialized with two covariance values  $\Sigma = [0.05, 0.1]$. 

Additionally, we include SVG-MPPI as a baseline, which represents the state-of-the-art MPPI-variant with mode-seeking behavior. We use the standard implementation and hyperparameters of SVG-MPPI~\cite{honda2024steinvariationalguidedmodel}.

\textbf{System Specs:} All simulation experiments are conducted on a Ubuntu 24.04 platform. The computer is equipped with an Intel i9-13900HX and a Nvidia GeForce RTX 4090. Real-world experiments were conducted on the F1Tenth racer platform~\cite{o2020f1tenth}, which runs ROS2 Foxy on Ubuntu 20.04 and is equipped with a Nvidia Jetson Xavier. The parameters for the vehicle are taken from~\cite{gonultas2023identificationcontrolfrontsteeredackermann}.

\textbf{Cost Function:} The cost function $J$ has two components: the state obstacle cost $\mathcal{C}_{\text{obs}}(x_t)$, which penalizes states based on the local costmap values to avoid obstacles and the distance-to-goal cost $\mathcal{C}_{\text{goal}}(x_t, x_{\text{goal}})$, which encourages the trajectory to minimize the distance to the goal $x_{\text{goal}}$. The total cost $J$ is computed over a time horizon $T$, and the relative importance of obstacle avoidance and goal-reaching is controlled by a weighting factor $\lambda$. Therefore, we have:

\vspace{-4pt}
\begin{equation}
\label{eq:cost_function}
    J^{\tau} = \phi(\statevec_T) +\sum_{t=0}^{T-1} \left( \lambda_{\text{obs}}\mathcal{C}_{\text{obs}}(\statevec_t^{\tau}) + \lambda_{\text{goal}} \mathcal{C}_{\text{goal}}(\statevec_t^{\tau}, \statevec_{\text{goal}}) \right)
\end{equation}
where a trajectory $\tau = F(\statevec_{\text{curr}},U) =\{\statevec_t^{\tau}\}_{t=0}^{T-1}$ and the initial state is equal to the current state $x_0^{T} = \statevec_{\text{curr}}$. The terminal cost is $\phi(\statevec_T) = \min_{t} \mathcal{C}_{\text{goal}}\left(\statevec_t^{\tau}, \statevec_{\text{goal}}) \right)$


To compute $\mathcal{C}_{\text{obs}}(x_t^{\tau})$, we define it based on the collision conditions:
\vspace{-3pt}
\begin{equation}
    \begin{split}
    &\mathcal{C}_{\text{obs}}(x_t^{\tau}) = \\
    &\begin{cases}
        \mathcal{C}_{\text{collision}}, & \text{if } \exists x_i^{\tau} \in \{x_i^{\tau}\}_{i=0}^{t-1} \text{ s.t. } x_i \text{ is in collision} \\
        \mathcal{C}_{\text{local}}(x_t^{\tau}), & \text{otherwise}
    \end{cases}
    \end{split}
\end{equation}
where $\mathcal{C}_{\text{collision}}$ is the max collision cost, and         $\mathcal{C}_{\text{local}}(x_t^{\tau})$ calculates the cost of robot footprint of the state based on the local costmap.

We define the goal cost function $\mathcal{C}_{\text{goal}}(x_t^{\tau}, x_{\text{goal}})$ as follows:
\vspace{-15pt}
\begin{align}
& \mathcal{C}_{\text{goal}}(x_t^{\tau}, x_{\text{G}}) = \notag \\
&\begin{cases}
    \mathcal{C}_{\text{distance}}, & \text{if } \exists x_i^{\tau} \in \{x_i^{\tau}\}_{i=0}^{t-1} \text{ s.t. } x_i \text{ is in collision} \\
    ||x_t^{\tau} - x_{\text{G}}||, & \text{otherwise}
\end{cases}
\end{align}
where $\mathcal{C}_{\text{distance}}$ is the goal cost of the state where the collision happened along a trajectory $\tau$. It is important to note that if any state in a trajectory $\tau$ reaches the goal, we stop the cost calculation. This means the trajectory cost is measured up to the goal-reaching state.

\subsection{Uniformity Analysis} \label{sec:uniformity_analysis_iros}
\begin{figure}[h]
    \includegraphics[width=\linewidth]{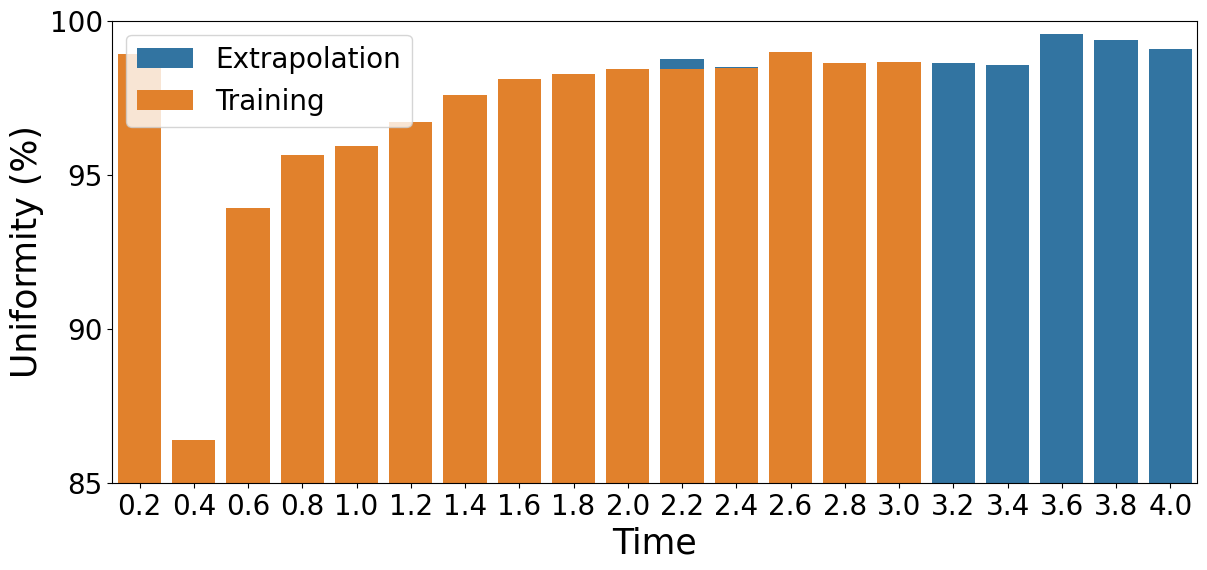}
    \caption{Uniformity Analysis:  \newourmethod{} learns to sample uniformly on all level sets. The X-axis represents the level sets for a 4-second time horizon with 0.2-second discretization. We test the uniformity performance of \newourmethod{} for extrapolation. The extrapolation (blue) experiment focuses on training with a 3-second time horizon with 0.2-second discretization and testing on a 4-second horizon which shows the capability of \newourmethod{} to plan for longer horizons. It can be observed that \newourmethod{} has high uniformity on all level sets.}
    \label{fig:dubins_uniformity_analysis}
\end{figure}

We investigate whether \newourmethod{} can sample each level set uniformly beyond the training distribution (extrapolation in time). To do so, we estimate the number of occurrences of each $\statevec_{D,t+1}$ in $L_{D,t+1}$ when sampled from $\statevec_{D,t}$ of $L_{D,t}$ using $p_\theta(\mathcal{U}|\statevec_{D,t})$. We then calculate the entropy with uniform samples of each $\statevec_{D,t+1}$ in $L_{D,t+1}$. The uniformity percentage metric is defined as the entropy ratio between occurrence distribution using \newourmethod{} and uniform distribution. We mainly perform the uniformity analysis for extrapolation in which the architecture is trained on level sets of 3 seconds with 0.2 discretization and is tested on 4 seconds 0.2 discretization. Fig. \ref{fig:dubins_uniformity_analysis} shows the uniformity percentage of \newourmethod{} on untrained level sets. It can be observed that \newourmethod{} is able to maintain high uniformity.

\vspace{-5pt}


\subsection{High Curvature Shortest Paths} \label{sec:open_space_exp_iros}
The experiments evaluate the performance of both baseline methods and our approach in configuration-to-configuration (C2C) navigation tasks within an open-space environment. We consider the same robot system as in Eq.~\ref{eq;kinematic_bicycle_model} over  4.5-second long trajectories. The initial state is fixed at $\statevec_{0} = [0,0,0]$.  
The cost function $J$ is the same as Eq.~\ref{eq:cost_function}, without the obstacle component. The cost weights are selected for each state element and the terminal cost as $\lambda_\statevec = [1.5,1.5,1.0]$ and $\lambda_{\phi} = 20.0$. We use $\cos\psi$, and $\sin\psi$ for the heading representation for the cost calculations. We design two sets of experiments. In the first experiment, the goal is to come back to the initial state which requires a circular motion --  the highest curvature maneuver. Second, we showcase the navigation performance for three challenging C2C tasks, each requiring a full turn. We run 10 trials for each method.

\begin{figure}[h]
    \centering
    \includegraphics[width=\linewidth, height = 5cm]{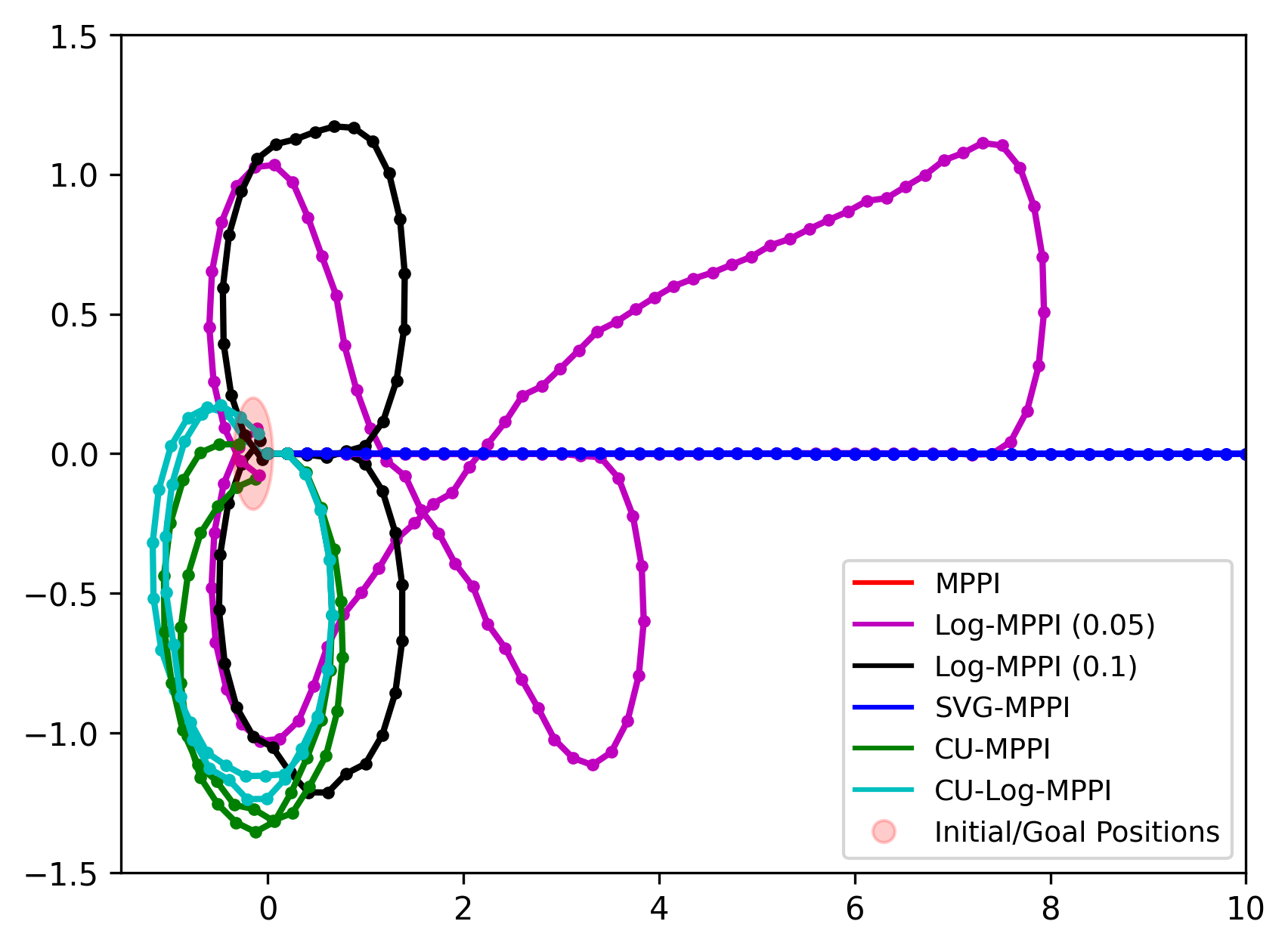}
    \caption{Circular Motion: \newsampourmethod{} (green), and \newsampourmethodlog{} (cyan), can navigate to a goal configuration while following the optimal path. The ability to generate high curvature turns by \newourmethod{} sampling helps in identifying the optimal regions. Log-MPPI (black and pink) achieves the closest results to our methods, but the higher variance (black) reduces the solution time. In contrast, MPPI (red) struggles due to limited exploration, and SVG-MPPI (dark blue) fails to steer trajectories effectively due to insufficient gradient information. Note that MPPI trajectories overlap with SVG-MPPI's trajectories, thus invisible in the figure. }
    \label{fig:full_turn_comp}
\end{figure}
\vspace{-3pt}
As shown in Fig.~\ref{fig:full_turn_comp},  \newourmethod{} generates high-curvature turns that help identify the low-cost regions, and navigates the vehicle towards the goal configuration. Among the baselines, log-MPPI achieves the highest success rates, while others have issues as a result of non-diversity in trajectories and ineffective gradient approximation that lead them to a failure.

\begin{figure*}[ht]
    \centering
    \begin{subfigure}{0.32\textwidth}
        \centering
        \includegraphics[width=\textwidth]{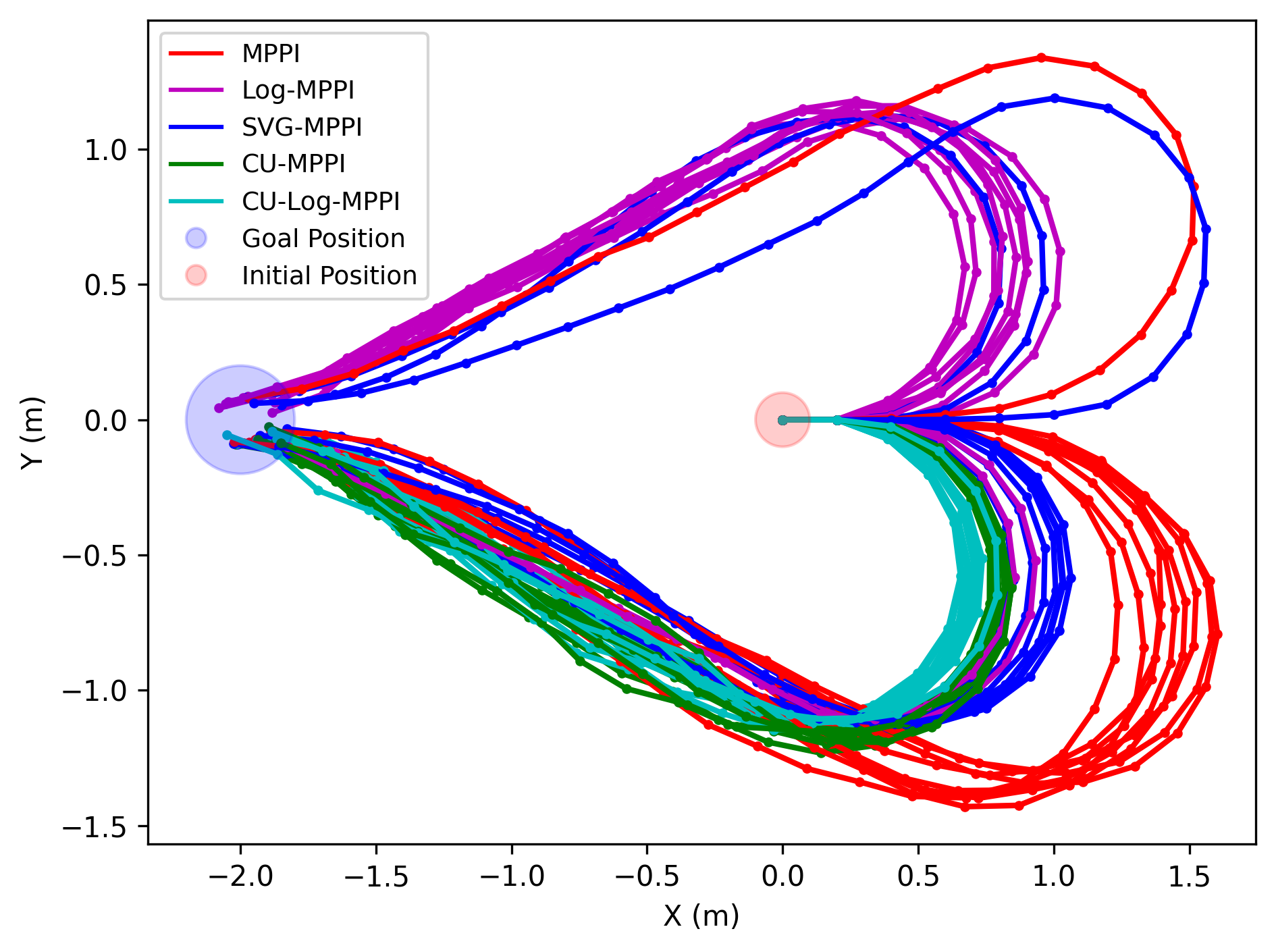}
        \caption{Target Configuration: $[-2,0,\pi]$}
        \label{fig:open_space_left}
    \end{subfigure}
    \hfill
    \begin{subfigure}{0.32\textwidth}
        \centering
        \includegraphics[width=\textwidth]{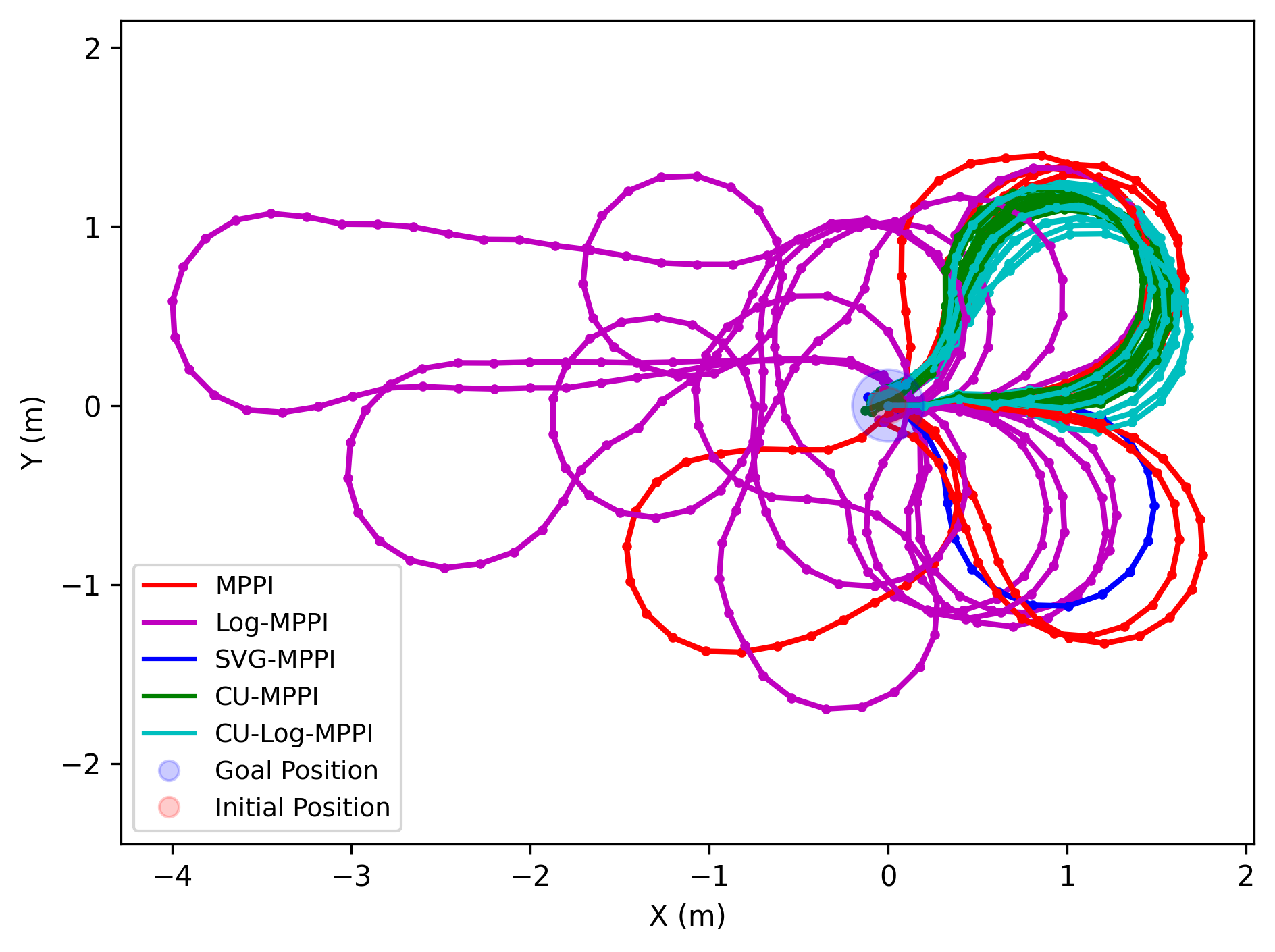}
        \caption{Target Configuration: $[0,0,\pi]$}
        \label{fig:open_space_center}
    \end{subfigure}
    \hfill
    \begin{subfigure}{0.32\textwidth}
        \centering
        \includegraphics[width=\textwidth]{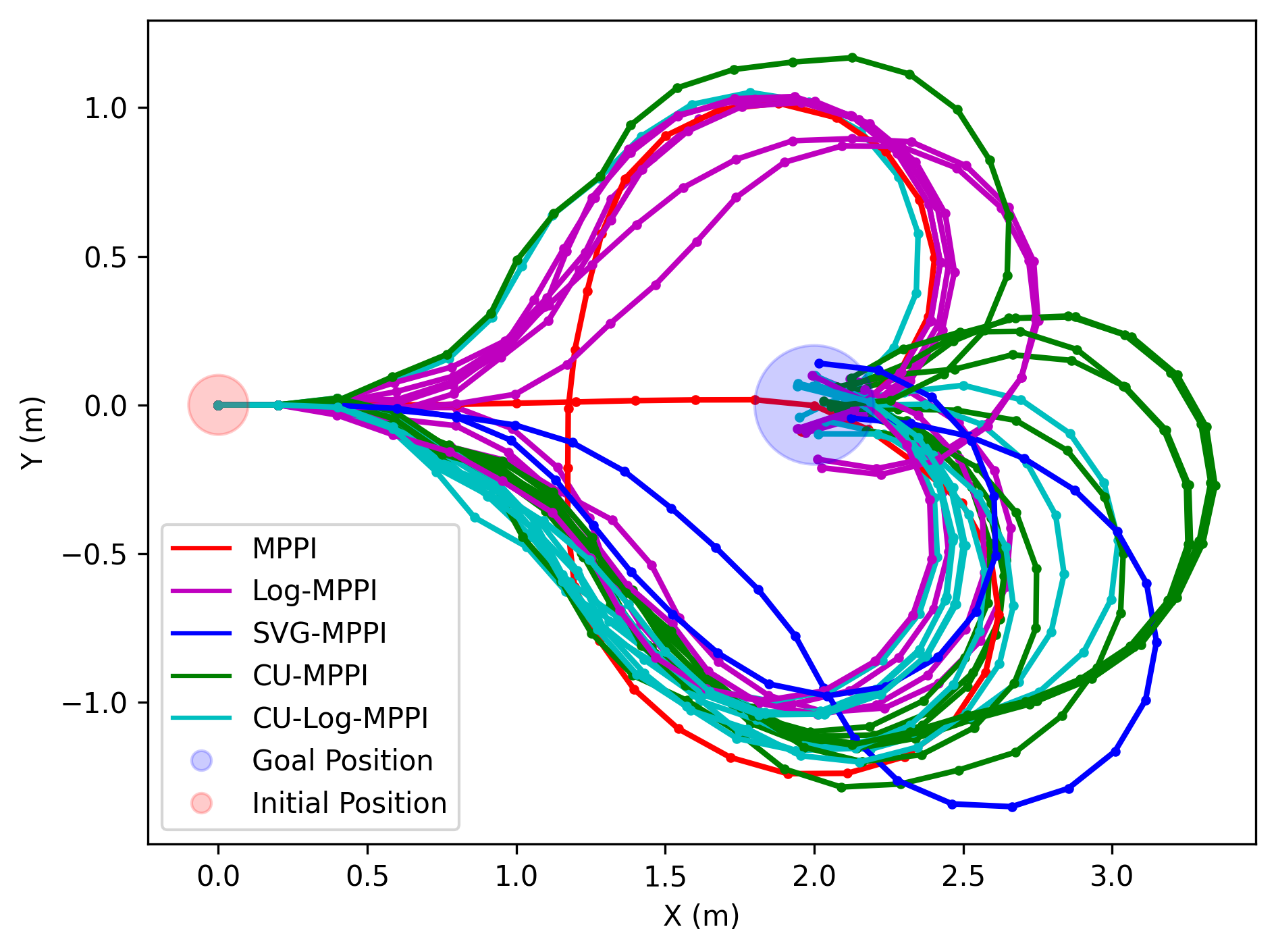}
        \caption{Target Configuration: $[2,0,\pi]$}
        \label{fig:open_space_right}
    \end{subfigure}
    \caption{Three  C2C tasks which demonstrate the performance of our method in high-curvature cases. The results highlight the number of successful runs out of 10 trials of the methods in these three settings. In (a), all methods reach the target configuration with CU-based methods and SVG-MPPI (dark blue) converging quicker than other baselines, which require more iterations to shift their distribution to the optimal one. In (b)-(c), \newsampourmethod{} (green), and \newsampourmethodlog{} (cyan) perform successfully whereas the other baselines start having issues to reach the target position due to the limited exploration for Log-MPPI (pink), MPPI (red), and insufficient gradient calculation for SVG-MPPI.}
    \label{fig:open_space_fig}
\end{figure*}

Fig.~\ref{fig:open_space_fig} highlights the differences in three more tasks: 
Fig.~\ref{fig:open_space_left} shows the adaptability of the approaches that shifts the nominal trajectory for sampling. It can be seen that SVG-MPPI and our methods can directly identify the optimal regions while the other two MPPIs need some iterations to find the low-cost areas. However, when the gradient information gets lost due to the complexity of the cost landscape, SVG-MPPI method starts to struggle. Similarly, when the need for high-curvature increases, the performance of the MPPI and log-MPPI decreases. 

\subsection{Dynamic Environment Simulation Experiments} \label{sec:simulation_exp_iros}
We investigate the navigation performance in complex and dynamic settings on a set of simulation environments by sudden obstacles appearing at varying distances from the vehicle. We design a set of cluttered environments with predetermined positions of 10, 15, 20, 25, and 30 circular obstacles of radius 1m in an environment size of $35m \times 10m$, resulting in a total of 50 environments. The difficulty is defined by increasing the number of obstacles and simulating the obstacles as sudden appearances at reducing distances from the vehicle. In particular, the vehicle has a constant egocentric detection range of $3m \times 3m$ but the obstacle is only revealed to the vehicle if it is within the experiment distance threshold:  1.5m, 1.25m, 1m, and 0.5m.  The shorter distances require more agility to react dynamically to obstacle appearing at different distances. Additionally, we also add small state noise to simulate localization error.

We evaluate the baselines on the dataset for success rate where success is defined when the vehicle reaches from initial location to goal location without any collisions. We choose one combination of initial and goal location for all the baselines and also compare the performance for different variance and number of trajectories. Table \ref{tab:comparison_success_dynamic} shows the comparison of the average success rate across all environments and distance thresholds of our methods \newsampourmethod{} and \newsampourmethodlog{} to the baselines. We also report the success rate when the obstacle distance threshold is the lowest which is 0.5m to show the ability of the approaches to rapidly evade the obstacle.
\begin{table}[ht]
    \caption{\small Dynamic Environment Success Rate ($\uparrow$) Performance Comparison with MPPI variants. Avg means average success across all environments and combinations while Obs Dist = 0.5m means when experiment distance threshold is 0.5m}
    \resizebox{\columnwidth}{!}{%
    \begin{tabular}{c|c|c|c|c|c|}
    \toprule
    \multicolumn{2}{c|}{\textbf{Num. of Traj.}}  & 
    \multicolumn{2}{c|}{\textbf{5000}} &
    \multicolumn{2}{c|}{\textbf{2500}} \\
    \midrule 
        \textbf{Methods} & \textbf{$\Sigma$} 
        & \textbf{Avg} & \textbf{Obs Dist = 0.5m}  & \textbf{Avg} & \textbf{Obs Dist = 0.5m} \\
    \midrule
    \multirow{2}{*}{MPPI}  
    & 0.05 & 0.58 & 0.02  & 0.51            & 0.1  \\
    & 0.1  & 0.78 & 0.28  & 0.75           & 0.18   \\
    \midrule
    
    \multirow{2}{*}{Log-MPPI}  
    & 0.05 & 0.85 & \textbf{0.50} & 0.84   & 0.48     \\
    & 0.1  & 0.86 & 0.50 & \textbf{0.89}   & \textbf{0.58}   \\ 
    \midrule
    
    \multirow{2}{*}{SVG-MPPI} 
    & 0.05 & 0.79 & 0.46  & 0.77           & 0.48   \\
    & 0.1  & 0.80 & 0.44  & 0.82           & 0.32  \\
    \midrule

    \multirow{2}{*}{\newsampourmethod{} (Ours)}  
    & 0.05 & 0.82 & 0.46 & 0.84     & 0.48   \\
    & 0.1  & 0.86 & 0.48  & \textbf{0.89}    & \textbf{0.58}    \\
    \midrule

    \multirow{2}{*}{\newsampourmethodlog{} (Ours)}  
    & 0.05 & \textbf{0.86} & \textbf{0.50}  & \textbf{0.88}    & \textbf{0.58}    \\
    & 0.1  & \textbf{0.87} & \textbf{0.52}   & 0.86   & 0.56    \\
    \bottomrule

    \end{tabular}%
    }
    \label{tab:comparison_success_dynamic}
\end{table}

\begin{figure}[!h]
    \includegraphics[width=\linewidth]{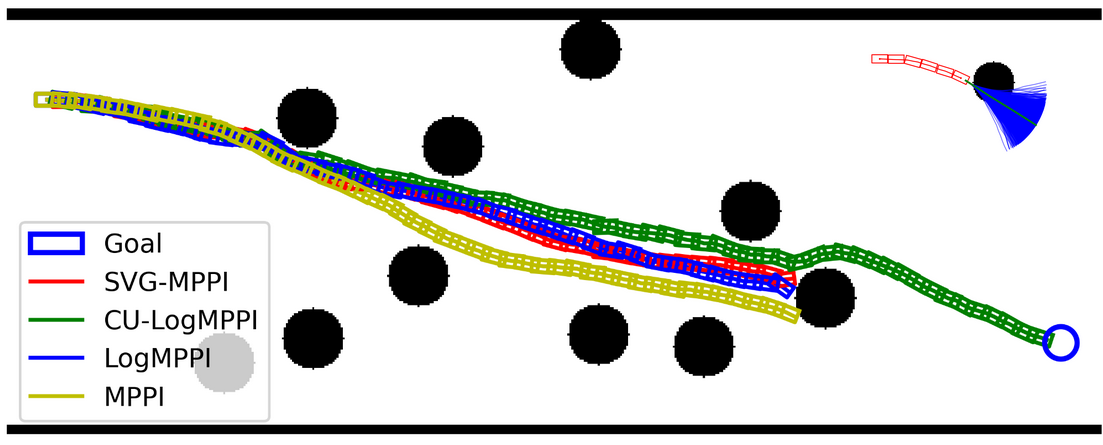}
    \caption{Low Variance with 2500 trajectories Obstacle Avoidance: The avoidance maneuver taken by different approaches when an obstacle suddenly appears in front of the vehicle. It can be seen that \newsampourmethodlog{} (green) performs the evading maneuver to avoid the obstacle while others fail. The top right shows sampled trajectories of MPPI in blue when a sudden obstacle appears in front of the vehicle where all trajectories lead to collision.}
    \label{fig:low_var_obs_avoid}
\end{figure}

It can be observed from Table \ref{tab:comparison_success_dynamic} that MPPI cannot perform evasive actions when sudden obstacles appear in front of the vehicle. This is because MPPI trajectories inherently do not have diversity. Hence with the appearance of sudden obstacles in its vicinity, all trajectories lead to collision. Another observation is that having an underlying trajectory distribution of C-Uniform improves the ability to evade sudden obstacles for low (0.05) variance and 2500 trajectories. Figure \ref{fig:low_var_obs_avoid} shows this scenario where it can be seen that having C-Uniform with Log-MPPI which is \newsampourmethodlog{} allows it to find the turn rapidly and evade the obstacle while other approaches fail as they do not have the diversity of trajectories from C-Uniform. This capability is useful for practical scenarios where vehicles often lack the computational capacity to process a high number of trajectories (5000) and using a high variance (0.1) tends to induce oscillations making it less suitable for real-world applications. Lastly, in the remaining scenarios all approaches perform similarly well.
\vspace{-5pt}

\subsection{Cluttered Environments}\label{sec:real_world_exp_iros}
We perform real-world and simulation experiments in cluttered environments. The 
BARN dataset~\cite{barn} is used to assess the performance of methods in both experiments. The simulation experiments were performed on the full BARN dataset, and the environments of real-world experiments are selected by separating the dataset into three difficulty groups and picking an environment from each group uniformly at random. These three environments are: Map 21 (easy) has relatively open areas, Map 110 (medium) has moderate clutter, and Map 289 (hard) has dense obstacles and narrow corridors. These maps provide a structured benchmark for assessing real-world navigation performance under varying difficulty levels. 
For these experiments, start and goal positions are fixed for all environments. Robot localization is performed using the Nav2 stack that employs the adaptive Monte Carlo localization method against a pre-built map, without pre-specified obstacle positions. The sensing range is clipped to a radius of 3 meters around the LiDAR sensor.  Each controller setting was evaluated over 10 trials and controlled at a frequency of 10 Hz. SVG-MPPI is excluded from the real-world experiments because it cannot run at 10 Hz on Nvidia Jetson Xavier due to hardware limitations. In simulation, SVG-MPPI achieved 0.85 and 0.87 success rates for variance setting 0.05 and 0.1 respectively.

\begin{table}[ht]
    \caption{\small Performance Comparison of Real-World and Simulation Experiments on Easy~(E), Medium~(M), and Hard~(H) Cluttered Environments. The average trajectory length is computed using only the successful trials.}
    \resizebox{\columnwidth}{!}{%
    \begin{tabular}{c|c|c|c|c|c|c|c|c|}
    \toprule
    \multicolumn{2}{c|}{}  & 
    \multicolumn{3}{c|}{\textbf{Real SR($\uparrow$)}} &
    \multicolumn{3}{c|}{\textbf{Real Avg. Length($\downarrow$)}} &
    \multicolumn{1}{c|}{\textbf{Sim SR($\uparrow$)}} \\
    \midrule 
        \textbf{Methods} & \textbf{$\Sigma$} 
        & \textbf{E} & \textbf{M} & \textbf{H} & \textbf{E} & \textbf{M} & \textbf{H} & \textbf{All}\\
    \midrule
    \multirow{2}{*}{MPPI}  
    & 0.05 & 0.8 & 0.4  & 0.1           & \textbf{12.2}  & \textbf{12.5}  & \textbf{12.5}  &0.86\\
    & 0.1  & 0.9 & 0.4  & 0.5           & \textbf{12.2}  & \textbf{12.5}  & 12.6  &0.92\\
    \midrule
    
    \multirow{2}{*}{Log-MPPI}  
    & 0.05 & \textbf{1.0} & 0.5 & 0.4   & \textbf{12.2}  & 17.1  & 12.9   &0.94\\
    & 0.1  & \textbf{1.0} & 0.7 & 0.4   & 12.4  & 12.9  & 12.9  &0.98\\ 
    \midrule
    

    \multirow{2}{*}{\newsampourmethod{} (Ours)}  
    & 0.05 & \textbf{1.0} & \textbf{0.9} & 0.6     & 12.7  & 12.8 & 13.0  &\textbf{1.0}\\
    & 0.1  & \textbf{1.0} & 0.8  & 0.8             & 12.6  & 12.9 & 13.2  &\textbf{1.0}\\
    \midrule

    \multirow{2}{*}{\newsampourmethodlog{} (Ours)}  
    & 0.05 & \textbf{1.0} & \textbf{0.9}  & \textbf{0.9}    & 12.5  & 12.8 & 13.0  &\textbf{1.0}\\
    & 0.1  & \textbf{1.0} & 0.8    & 0.8                    & 12.5  & 12.8 & 13.0  &\textbf{1.0}\\

    \bottomrule

    \end{tabular}%
        }
    \label{tab:comparison_real_world}
\end{table}

\begin{figure}[ht]
    \centering
    \includegraphics[width=0.73\linewidth, height = 10 cm]{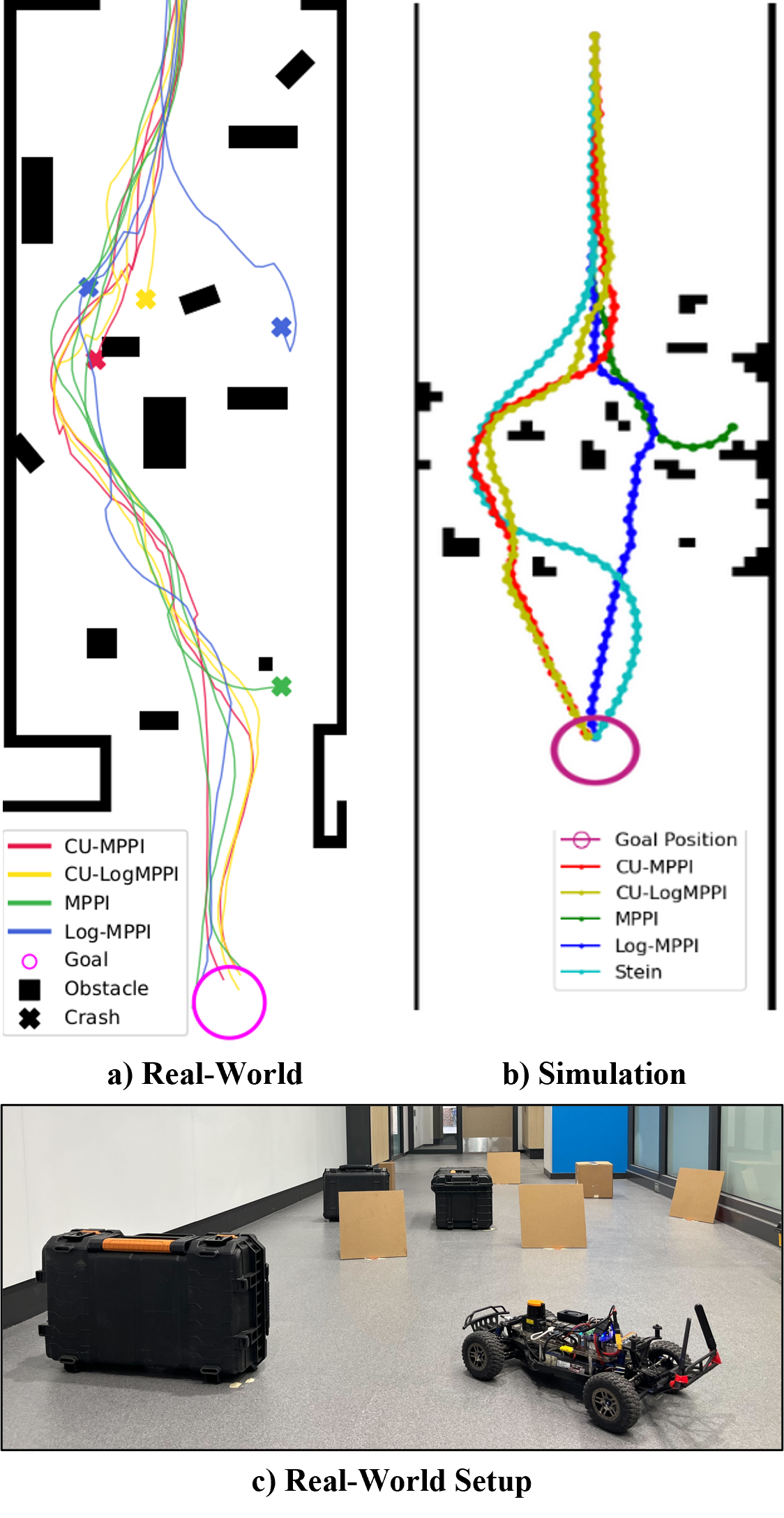}
    \caption{Real vs Sim: The trajectories in (a) are from the real-world scenario. The ones in (b) are from a similar environment (the ``hard" environment) in simulation. It can be observed that the trajectories taken in the simulation are very similar to those of the real-world. Additionally, in (c), we show the vehicle and the real-world setup.}
    \label{fig:real_world_exp}
\end{figure}

 Table \ref{tab:comparison_real_world} shows the success rate reported and the average trajectory length among successful runs using 1500 trajectories for both simulation and real-world. Log-MPPI with a variance of 0.05 has a looping trajectory in the medium-difficulty environment, which increases the average trajectory length dramatically. \newsampourmethod{} and \newsampourmethodlog{} show higher success rates in medium and high-difficulty environments compared to baselines while maintaining a similar average trajectory length to Log-MPPI. In simulation, \newsampourmethod{} and \newsampourmethodlog{} outperform all the approaches by achieving 100\% collision-free paths for all environments. The improved performance of all the approaches in simulation is directly related to the localization and hardware noise in real-world experiments. As an additional validation of the real-world applicability of our results, Figure \ref{fig:real_world_exp} shows a side-by-side comparison of trajectories generated in real and simulation for the same environment.
 


\vspace{-5pt}
\section{Conclusion}
In this work, we presented a new approach to choose control input probabilities to sample trajectories which are C-Uniform: At each time step $t$, and for each subset $S$ of the level set $L_t$, the probability that the robot is in $S$ is proportional to the measure of $S$. In contrast to our previous work~\cite{poyrazoglu2024cuniformtrajectorysamplingfast} in which the probabilities are obtained by building a flow network based on a discretization of the configuration space, our new approach is based on learning the weights of a neural network which maps robot states to action probability distributions using entropy as unsupervised loss. It mitigates scalability issues of our previous approach in terms of both spatial resolution and time-horizon.  Next, we showed how the C-Uniform trajectory sampler can be coupled with a local sampling-based gradient-follower to obtain a novel MPPI variant, \newsampourmethod. Our method outperforms existing MPPI variants in high curvature settings. 

Our current implementation of \newsampourmethod{} uses a pre-built map of the environment for localization (the obstacles are not necessarily pre-mapped). In our future work, we are planning to incorporate localization into navigation to remove this dependency.

\label{sec:conclusion}



\bibliographystyle{unsrt}
\bibliography{root}

\begin{thebibliography}{10}

\bibitem{poyrazoglu2024cuniformtrajectorysamplingfast}
O.~Goktug Poyrazoglu, Yukang Cao, and Volkan Isler.
\newblock {C-Uniform} trajectory sampling for fast motion planning, 2024.
\newblock accepted to ICRA 2025.

\bibitem{williams2018information}
Grady Williams, Paul Drews, Brian Goldfain, James~M Rehg, and Evangelos~A Theodorou.
\newblock Information-theoretic model predictive control: Theory and applications to autonomous driving.
\newblock {\em IEEE Transactions on Robotics}, 34(6):1603--1622, 2018.

\bibitem{lambert2020stein}
Alexander Lambert, Adam Fishman, Dieter Fox, Byron Boots, and Fabio Ramos.
\newblock Stein variational model predictive control.
\newblock {\em arXiv preprint arXiv:2011.07641}, 2020.

\bibitem{minavrik2024model}
Michal Mina{\v{r}}{\'\i}k, Robert P{\v{e}}ni{\v{c}}ka, Vojt{\v{e}}ch Von{\'a}sek, and Martin Saska.
\newblock Model predictive path integral control for agile unmanned aerial vehicles.
\newblock In {\em 2024 IEEE/RSJ International Conference on Intelligent Robots and Systems (IROS)}, pages 13144--13151. IEEE, 2024.

\bibitem{mohamed_autonomous_2022}
Ihab~S Mohamed, Kai Yin, and Lantao Liu.
\newblock Autonomous navigation of agvs in unknown cluttered environments: log-mppi control strategy.
\newblock {\em IEEE Robotics and Automation Letters}, 7(4):10240--10247, 2022.

\bibitem{honda2024steinvariationalguidedmodel}
Kohei Honda, Naoki Akai, Kosuke Suzuki, Mizuho Aoki, Hirotaka Hosogaya, Hiroyuki Okuda, and Tatsuya Suzuki.
\newblock Stein variational guided model predictive path integral control: Proposal and experiments with fast maneuvering vehicles, 2024.

\bibitem{kalakrishnan2011stomp}
Mrinal Kalakrishnan, Sachin Chitta, Evangelos Theodorou, Peter Pastor, and Stefan Schaal.
\newblock Stomp: Stochastic trajectory optimization for motion planning.
\newblock In {\em 2011 IEEE international conference on robotics and automation}, pages 4569--4574. IEEE, 2011.

\bibitem{kazim2024recent}
Muhammad Kazim, JunGee Hong, Min-Gyeom Kim, and Kwang-Ki~K Kim.
\newblock Recent advances in path integral control for trajectory optimization: An overview in theoretical and algorithmic perspectives.
\newblock {\em Annual Reviews in Control}, 57:100931, 2024.

\bibitem{orthey_sampling-based_2024}
Andreas Orthey, Constantinos Chamzas, and Lydia~E Kavraki.
\newblock Sampling-based motion planning: A comparative review.
\newblock {\em Annual Review of Control, Robotics, and Autonomous Systems}, 7, 2023.

\bibitem{ota_trajectory_2019}
Kei Ota, Devesh~K Jha, Tomoaki Oiki, Mamoru Miura, Takashi Nammoto, Daniel Nikovski, and Toshisada Mariyama.
\newblock Trajectory optimization for unknown constrained systems using reinforcement learning.
\newblock In {\em 2019 IEEE/RSJ international conference on intelligent robots and systems (IROS)}, pages 3487--3494. IEEE, 2019.

\bibitem{lavalle_planning_2006}
Steven~M LaValle.
\newblock {\em Planning algorithms}.
\newblock Cambridge university press, 2006.

\bibitem{power2024learning}
Thomas Power and Dmitry Berenson.
\newblock Learning a generalizable trajectory sampling distribution for model predictive control.
\newblock {\em IEEE Transactions on Robotics}, 2024.

\bibitem{abughalieh2019survey}
Karam~M Abughalieh and Shadi~G Alawneh.
\newblock A survey of parallel implementations for model predictive control.
\newblock {\em IEEE Access}, 7:34348--34360, 2019.

\bibitem{yin2022trajectory}
Ji~Yin, Zhiyuan Zhang, Evangelos Theodorou, and Panagiotis Tsiotras.
\newblock Trajectory distribution control for model predictive path integral control using covariance steering.
\newblock In {\em 2022 International Conference on Robotics and Automation (ICRA)}, pages 1478--1484. IEEE, 2022.

\bibitem{trevisan2024biased}
Elia Trevisan and Javier Alonso-Mora.
\newblock Biased-mppi: Informing sampling-based model predictive control by fusing ancillary controllers.
\newblock {\em IEEE Robotics and Automation Letters}, 2024.

\bibitem{asmar2023model}
Dylan~M Asmar, Ransalu Senanayake, Shawn Manuel, and Mykel~J Kochenderfer.
\newblock Model predictive optimized path integral strategies.
\newblock In {\em 2023 IEEE International Conference on Robotics and Automation (ICRA)}, pages 3182--3188. IEEE, 2023.

\bibitem{kobayashi2022real}
Taisuke Kobayashi and Kota Fukumoto.
\newblock Real-time sampling-based model predictive control based on reverse kullback-leibler divergence and its adaptive acceleration.
\newblock {\em arXiv preprint arXiv:2212.04298}, 2022.

\bibitem{guiasu1985principle}
Silviu Guiasu and Abe Shenitzer.
\newblock The principle of maximum entropy.
\newblock {\em The mathematical intelligencer}, 7:42--48, 1985.

\bibitem{o2020f1tenth}
Matthew O'Kelly, Hongrui Zheng, Dhruv Karthik, and Rahul Mangharam.
\newblock F1tenth: An open-source evaluation environment for continuous control and reinforcement learning.
\newblock {\em Proceedings of Machine Learning Research}, 123, 2020.

\bibitem{gonultas2023identificationcontrolfrontsteeredackermann}
Burak~M. Gonultas, Pratik Mukherjee, O.~Goktug Poyrazoglu, and Volkan Isler.
\newblock System identification and control of front-steered ackermann vehicles through differentiable physics, 2023.

\bibitem{barn}
Daniel Perille, Abigail Truong, Xuesu Xiao, and Peter Stone.
\newblock Benchmarking metric ground navigation.
\newblock In {\em 2020 IEEE International Symposium on Safety, Security, and Rescue Robotics (SSRR)}, pages 116--121, 2020.

\end{thebibliography}

\end{document}